\documentclass[10pt]{article}
\usepackage{graphicx}
\usepackage{url}
\begin{document}
\begin{center}
    {\large\bf Vector potential, electromagnetic induction and ``physical meaning''}\\
    \vskip3mm
    Giuseppe Giuliani\\
    \vskip3mm
    {\small Dipartimento di Fisica ``Volta'', Universit\`a degli Studi di Pavia, via Bassi 6, 27100 Pavia, Italy}\\
    {\small Email: {\verb"giuseppe.giuliani@unipv.it"}}
\end{center}
\thispagestyle{empty}
\small
{\bf Abstract.}
A forgotten experiment by Andr\'e  Blondel (1914) proves, as held on the basis of theoretical arguments in a previous paper, that the time variation of the magnetic flux is not the cause of the induced $emf$:  the physical agent is instead the vector potential through the term $-\partial\vec A/\partial t$ (when the induced circuit is at rest).
 The ``good electromagnetic potentials'' are determined by the Lorenz condition and  retarded formulas.
 Other pairs of potentials derived through appropriate gauge functions are only mathematical devices for calculating the fields: they are not physically related to the sources.
 The physical meaning of a theoretical term relies, primarily, on theoretical grounds: a theoretical term has physical meaning if it cannot be withdrawn without reducing the predictive power of a theory or, in a weaker sense, if it cannot be withdrawn without reducing the descriptive proficiency of a theory.
\normalsize
\section{Introduction}
The history of the role of vector potential~-~and, more generally, of potentials~-~in electromagnetism is an intricate one. A reader willing to go through the vast literature might consider the compendious paper by     Roche as a starting point \cite{roche}, the paper by Jackson and Okun as a second step \cite{jack} and the book by Darrigol \cite{darri} for an overall picture of nineteenth century electrodynamics.
\par
 The main thread may be described as follows: does the vector potential have a ``physical meaning''\footnote{We shall discuss the meaning of ``physical meaning'' in section \ref{what}.} or is it just a mathematical device that could, in the end, be dismissed? As the historical development shows, the issue has physical and epistemological relevance.
Therefore, it cannot be neglected in teaching electromagnetism to physics students at university (undergraduate/graduate) level.  In particular, teachers should be well aware of the subtleties underlying the interplay between fields and potentials.
\par
It is well known that Maxwell  made ample use of a ``physically meaningful''  vector potential in the description of electromagnetic phenomena \cite{roche} \cite{jack} \cite{darri} \cite{rou}.
However, at the end of the nineteenth century, owing to the challenges by Heaviside and Hertz \cite{roche} \cite{jack} \cite{darri}, the vector potential was reduced to a ``magnitude which serves as calculation only''.
Heaviside:
``The method by which Maxwell deduced \dots [the  $emf$ induced in a conductor by its motion  in a magnetic field] is substantially the same in principle; he, however, makes use of an auxiliary function, the vector~-~potential of the electric current, and this rather complicates the matter, especially as regards the physical meaning of the process. It is always desirable when possible to keep as near as one can to first principles \cite[page 46]{heavi}.''
And Hertz: ``In the construction of the new theory the potential served as a scaffolding; by its introduction the distance-forces which appeared discontinuously at particular point were replaced by magnitudes which at every point in space were determined only by the condition at the neighbouring points. But after we have learnt to regard the forces themselves as magnitudes of the latter kind, there is no object in replacing them by potentials unless a mathematical advantage is thereby gained. And it does not appear to me that any such advantage is attained by the introduction of the vector-potential in the fundamental equations; furthermore, one would expect to find in these equations relations between the physical magnitudes which are actually observed, and not between magnitudes which serves for calculations only \cite[page 195]{hertzvp}.''
\par
Thereafter, according to \cite{roche}, the vector potential has been usually treated in textbooks according to Heaviside's and Hertz's approach  until the appearance of Feynman {\em Lectures} with his discussion about what must be considered as a ``real field''.
According to Feynman, a ``real field'' is ``a mathematical function we are using for avoiding the idea of action at a distance''  \cite[page 15.7]{fyepv}. This definition appears as an unwitting  quotation of Hertz's comment reported above: however, Feynman does not reach Hertz's conclusion that the vector potential can be dismissed. In fact, Feynman writes that ``We have introduced $\vec A$ because it {\em does} have an important physical significance \dots In any region where $\vec B=0$ even if $\vec A$ is not zero, such as outside a solenoid, there is no discernible effect of $\vec A$. Therefore, for a long time it was believed that $\vec A$ was not a ``real'' field. It turns out, however, that there are phenomena involving quantum mechanics [Aharonov~-~Bohm effect] which show that the field $\vec A$ is in fact a ``real'' field in the sense we have defined it \cite[page 15.8]{fyepv}.''
Clearly, according to Feynman,  the vector potential acquires  ``physical meaning''  only from quantum  phenomena.
However, more recently, the
     ``physical meaning'' of vector potential in classical electromagnetism has been advocated in connection with a particular phenomenon of electromagnetic induction: the so called Maxwell~-~Lodge effect   \cite{rou} \cite{matteucci}. In these papers, the Maxwell~-~Lodge effect is considered~-~correctly as we shall see~-~as the classical counterpart of the Aharonov~-~Bohm effect (section \ref{outside}).
  \par
      In this paper
       we shall:
        \begin{enumerate}
          \item comment on ``a general law for electromagnetic induction'' \cite{epl} based on an {\em essential} use of the vector potential (section \ref{glawsec});
          \item discuss a forgotten experiment by Andr\'e Blondel (1914) \cite{blondel}  which shows that the variation of the magnetic flux through a surface that has a conducting  circuit as a contour is not the cause of the $emf$ induced in the circuit (section \ref{blonsec});
          \item reflect on the infinite pairs of potentials (scalar and vector potential) usually considered as physically equivalent because yielding the same electric and magnetic fields (section \ref{final});
              \item discuss the meaning of ``physical meaning'' (section \ref{what}).
        \end{enumerate}
     As the above list shows, this paper tries to shed new light on some basic  topics of classical electromagnetism and on the long debated issue of ``physical meaning''. Therefore, it may be of some usefulness to university teachers and  graduate students.

                       \section{A general law for electromagnetic induction\label{glawsec}}
In a recent paper \cite{epl} a general law for electromagnetic induction phenomena has been derived by defining the induced $emf$ as the integral, over a closed path, of the Lorentz force on the unit positive charge:
\begin{equation}  \label{forzaem} {\cal E}= \oint
_{l}^{}{(\vec E + \vec v_{c} \times \vec
B)\,\cdot\,\vec{dl}}
\end{equation}
By putting:
\begin{equation}\label{vecpot}
    \vec E = - grad\, \varphi -{{\partial \vec A}\over{\partial t}}
\end{equation}
($\varphi$ scalar potential; $\vec A$ vector potential) one gets immediately the `general law' for electromagnetic induction:
\begin{equation}\label{leggegen}
    \mathcal E= -\oint_l \frac{\partial \vec A}{\partial t}\cdot \vec
dl
    +\oint_l (\vec v_{c}\times \vec B)\cdot\vec dl
\end{equation}
The
 two terms of this equation represent, respectively, the contribution  of the time variation of the vector potential and
  the effect of the magnetic field on  moving charges.
  Among the  main conclusions of \cite{epl}:
  \begin{enumerate}
       \item [(a)]
       The `flux rule' is not a physical law but a {\em calculation shortcut}: it yields the correct $emf$ when the induced circuit is filiform (or equivalent to) \footnote{With an exception to be discussed below.}. However, even in these cases,  the `flux rule' does not describe the physical processes responsible for  the observed phenomena.
           \item[(b)] \label{not}The variation of the flux of the magnetic field through an arbitrary surface
that has the circuit as contour {\em is not the cause of the
induced $emf$}.
\item[(c)] The general law (\ref{leggegen}) has been firstly written by Maxwell with an unavoidable, in those times, ambiguity about the meaning of the velocity appearing in it.
  \end{enumerate}
  \section{Blondel's experiment (1914)\label{blonsec}}
  A coworkers of mine,  Paolantonio Marazzini, has drawn my attention to a beautiful and forgotten experiment carried out  by Andr\'e Blondel \cite{blondel} about one century ago \footnote{For a biography of Andr\'e Blondel, see, for instance, \cite{blondel_bio}.  Around 1910 the electrical engineers community was debating about the validity of the flux rule: see, for instance the paper by Hering and the following discussion \cite{hering}.}.
   \begin{figure}[htb]
  \centering
  \includegraphics[width=4cm]{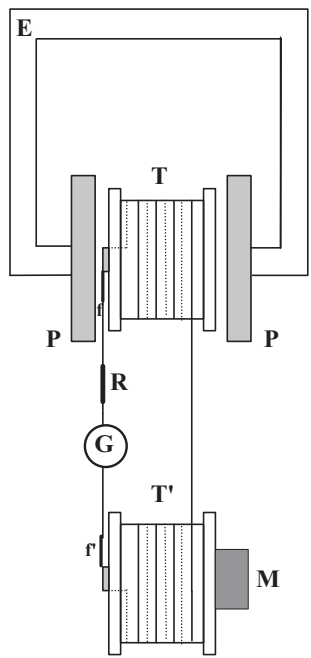}\\
  \caption{experimental setup used by Blondel. See text.}\label{blondel}
\end{figure}
\par\noindent
  The experimental set up  is shown in fig. \ref{blondel}.
$E$ is the core of an electromagnet that ends in the circular plates $PP$ (diameter $24\, cm$; $26\, cm$ apart); the magnetic field between the plates is ``sensibly uniform'' and equal to $0.08\, T$; $T$ is a wooden cylinder (diameter $20\, cm$) that can rotate about its axis; an isolated copper wire (section $0.5\,mm^2$) is winded on $T$ in several regular layers: one of its ends is fixed to the left of $T$ at a point specified in the text; the other end is fixed to the cylinder $T'$ (identical to $T$) in such a way that when $T'$ is connected to the motor $M$ already in motion, $T$ begins to rotate  by assuming ``very rapidly'' a constant angular velocity (thus moving windings from $T$ to $T'$); the fixed ends of the coil on $T$ and $T'$ are connected to the Arsonval galvanometer $G$ through the sliding contacts $f,f'$; $R$ is a resistance of $10000\, \Omega$. The sensibility of the galvanometer ``in the circuit so composed, produced a deviation of $24\, mm$ at [a distance of]  one meter in correspondence of  $ 0.005\,V$ ''. {\em The total number of windings $(N+N')$ on $T$ and $T'$, respectively, is constant: the synchronous rotation of the cylinders changes the number of windings inside the magnetic field thus producing a variable magnetic flux through the total number of windings $(N+N')$}.
\par
Blondel performed several experiments:
\begin{enumerate}
  \item [I.] The fixed end of the coil wound on $T$ is soldered to a thin copper ring fastened to the axis of $T$: the radius of the axis is negligible with respect to that of $T$ (this set up is illustrated in fig. \ref{blondel}). The measured $emf$, when the cylinder $T$ rotates at a speed of $6.66$ cycles per second, is of $0.015\,V$ \footnote{Blondel performed the experiments at a speed of $400$ cycles per minute ($6.66$ cycles per second). However, he gives the measured $emf$ ``normalized'' at ten cycles per second. We shall use Blondel ``true values'', i.e., those corresponding to the actual rotation speed.}.
  \item [II.] The fixed end of the coil on $T$ is soldered to a ring with the same diameter of $T$: the measured $emf$ is null.
  \item [III.] The fixed end of the coil on $T$ is connected, through a sliding contact, to the periphery of a {\em fixed} disc with the same diameter of $T$, parallel to it and with a hole through which passes the axis of $T$; the wire to the galvanometer is soldered to a point of the inner periphery of the disc. The measured $emf$ is null.
  \item [IV.]The fixed end of the coil on $T$ is directly soldered to the periphery of a disc (with the same diameter of $T$) rigidly connected to $T$, while a sliding contact connects the center of the disc with the galvanometer. The measured $emf$ is the same as in (I). If the sliding contact moves along a radius of the disc, then the measured $emf$ shows a parabolic dependence   on  the distance of the sliding contact from the center of the disc.
 \end{enumerate}
Blondel's experiments are easily described by the general law (\ref{leggegen}).
Since the vector potential is constant, the first term of (\ref{leggegen}) is null while the second term is different from zero only when the circuit
has a {\em rotating} part  along a radius of $T$ (I, IV)
\begin{quote}\small
\label{above}When the cylinders rotate each small piece of wire on $T$ moves with a velocity of about $2\times\pi\times  6.66 \times 0.1=4.18\,ms^{-1}$. This velocity is, with a good approximation, tangent to a circumference centered on and perpendicular to the axis of $T$: therefore, its contribution to the induced $emf$ through a term $\vec v\times\vec B$ is null. This is true also for the pieces of wire between $T$ and $T'$. As a matter of fact, there should be a small contribution to the induced $emf$ when the coil on $T$ passes from a layer to another (always due to a term $\vec v\times\vec B$): the several layers of wire winded on $T$ are equivalent to a piece of wire disposed along a radius of $T$ with a length equal to the thickness of the layers. According to Blondel, this contribution~-~of course not explained in terms of $\vec v\times\vec B$~-~is responsible for a ``small, apparently continuous, decrease'' of the measured $emf$ with the setups (I) and (IV). The order of magnitude of the effect due to a single passage between two layers (trough equation (\ref{disco}) below and assuming the diameter of the  wire  equal to $1\, mm$) is $\approx 3.3\times 10^{-4}\, V$. This value  corresponds to a galvanometer deviation   of about $1.5\, mm$ and should be compared with the total deviation of $70\, mm$.
\end{quote}
From our viewpoint, the most relevant result is  given by experiments (II) and (III). In these cases, according to the flux rule, the induced $emf$ should be:
\begin{equation}\label{fr}
    |\mathcal E|= \left|-\phi \frac{dN}{dt}\right|= \left|\phi \frac{dN'}{dt}\right|
\end{equation}
where $\phi$ is the magnetic flux through a single loop of $T$, $N$ the number of loops on $T$ and $N'$ the number of loops  on $T'$. If: $R=10\,cm$, $B=0.08\,T$ and $ |dN/dt|=6.66$, the $emf$ predicted by (\ref{fr}) will be equal to $0.017\,V$. This value is the same as the one predicted by equation (\ref{leggegen}) for experiments I and IV: in these experiments the measured $emf$ was of $0.015\, V$ corresponding to a galvanometer deviation of $70\, mm$. Since, in  experiments (II) and (III) the measured $emf$ is null (no galvanometer deviation), they {\em prove} that a variation of the magnetic flux does not produce an induced  $emf$ \footnote{As a matter of fact, also in this case, there should be an effect due to the passage between two adjacent layers: see the insert  above. Has Blondel not seen or neglected this small galvanometer deviation? }. These experiments corroborate the {\em theoretical} statement according to which
``The flux of the magnetic field through an arbitrary surface
that has the circuit as contour {\em is not the cause} of the
induced $emf$  \cite[page 60002~-~p6]{epl}.''
\par
This is a crucial point. Let us suppose, for a moment, that a time variation of the magnetic flux is the cause of the induced $emf$. The time variation of the flux can be due to a time variation of the magnetic field or/and to a time variation of the surface through which the flux is calculated. How can be sustained that the time variation of the flux produces an induced $emf$ only in the case of a varying magnetic field and not in the case of a varying surface?
If $A$ causes $B$, this should be true, {\em ceteris paribus}, independently of how $A$ is produced.
Therefore,  we must conclude that
 if the variation of the magnetic flux is a physical cause of something else, it should be so independently of how its variation is obtained.
    \par
  Blondel's experiments allow also a quantitative check of other  predictions of equation (\ref{leggegen}).
As a matter of fact:
\begin{itemize}
  \item  Since  case  (I) corresponds to a particular configuration of Faraday disc (discussed in \cite{epl}), the predicted $emf$ is given by:
  \begin{equation}\label{disco}
    \mathcal E= \frac{1}{2} B \omega R^2
  \end{equation}
  If: $R=10\,cm$, $B=0.08\,T$ and $\omega=2\times \pi\times 6.66$, we get $\mathcal E= 0.017\, V$. The measured value is a little lower ($0.015\, V$),  due to the fact that the radius of the axis of $T$ is not null.
  \item  In case (IV) the predicted value is the same as for (I) when the sliding contact is at the center of the disc. When the sliding contact is at a distance $d$ from the center, equation (\ref{leggegen}) predicts
      \begin{equation}\label{casod}
        \mathcal E= \frac{1}{2}B\omega (R^2 -d^2)
      \end{equation}
      i.e. a parabolic dependence on $d$, as observed by Blondel.
  \end{itemize}
\section{\label{final}Which  potentials? }
Taken for granted the role of  vector potential in electromagnetic induction phenomena through the quantity $-\partial \vec A/\partial t$, we shall now focus on  the fact that it is possible to construe an infinite numbers of couples of electromagnetic potentials that yield the same fields. If $\vec A$ and $\varphi$ are ``good'' potentials, then also $\vec A'=\vec A+\nabla \psi$ and $\varphi'=\varphi-\partial \psi/\partial t$ are good potentials because they yield the same fields through the relations $\vec B= \nabla\times\vec A$ and $\vec E= -\nabla\, \varphi- \partial \vec A/\partial t$ (and similar relations for the primed quantities): $\psi$ is called gauge function.
However,  the question: ``how can we get good starting potentials?'' is often overlooked. The  answer  is given by the following
procedure, {\em dictated} by the  electromagnetic theory:
\begin{enumerate}
  \item [(P1)]  Maxwell equations for the fields contain the continuity equation between the sources $\nabla\cdot \vec J= -\partial\rho/\partial t$; consequently, a continuity equation must  hold also for the potentials. This equation {\em must} be Lorentz invariant: therefore, it must be given by the so called Lorenz condition:  $\nabla\cdot \vec A =- \varepsilon_0\mu_0 (\partial\varphi/\partial t)$.
    \item [(P2)] If condition (P1) is satisfied, then the  potentials are given by the ``retarded'' formulas
    \begin{equation}
\!\!\!\!\!\varphi (x_1,y_1,z_1,t)  =  {{1} \over {4\pi \varepsilon_0 }}
\int_{}^{}{{{\rho(x_2,y_2,z_2, t-r_{21}/c)} \over {r_{21}}} } \:
d\tau_2 \label{scalaB}
\end{equation}
\begin{equation}
\!\!\!\!\! A_i (x_1,y_1,z_1,t) =  {{\mu_0} \over {4\pi}} \int_{}^{}{{{
J_i(x_2,y_2,z_2, t-r_{21}/c) } \over {r_{21}}} } \: d\tau_2
 \label{potenziaB}
\end{equation}
    The sources $\vec J$ and $\rho$ are enclosed in a finite volume $\tau_2$ so that $\vec A$ and $\varphi$ go to zero as the distance from the sources goes to $\infty$.
              \end{enumerate}
This procedure determines in a unique way the potentials; any other couple of potentials obtained by an appropriate gauge function are only (unnecessary) mathematical expressions which allow the calculation of the fields but are not physically related to the sources.
\par
For further clarifying this point, let us apply this procedure to the well
 known example  of the vector potential of a uniform magnetic field. It is usual to read in textbooks that the vector potential of a uniform magnetic field directed along the $z$ direction is \cite[page 14.2]{fyepv}:
 \begin{equation}\label{potcirc}
    \frac{1}{2}\vec B\times \vec r
    \qquad A_x= -\frac{1}{2} B_z y \:\:A_y=\frac{1}{2}B_z x\:\:A_z=0
\end{equation}
or, alternatively
 \begin{eqnarray}
    A_x=0,\,A_y=B_zx,\,A_z&=&0\\
   \rm{or}&&\label{landau1}\nonumber\\
    A_x=-B_zy,\,A_y=0,\,A_z&=&0
 \end{eqnarray}
However,  when speaking of a uniform magnetic field, we should specify which are the sources of the field. In the following, we shall deal with an ideal (indefinite) linear solenoid in which a  slowly varying current flows.
In this case, since $\rho=0$ everywhere, the scalar potential may be assumed to be zero. Then, rule (P1) (the Lorenz condition)  implies $\nabla\cdot\vec A=0$. As a matter of fact, we are in  the ``magnetic galilean limit" of relativistic electromagnetism \cite{jml} \cite{rou2}. Both expressions (\ref{potcirc}) and (\ref{landau1}) satisfy this condition.
Before applying rule (P2), we must specify the approximation we are working with.
Since $r_{21} \sim 1\, m$, the retardation time  $\sim 0.33 \times 10^{-8}\, s$, can be neglected in (\ref{scalaB}, \ref{potenziaB}): therefore, the relations between sources and potentials become instantaneous. Furthermore, we suppose that, in case of sinusoidal current, its frequency is low enough for neglecting wave irradiation.
Owing to the cylindrical symmetry of the problem, the vector potential due to the current in the solenoid must be a vector tangent to a circumference centered on and perpendicular to the solenoid axis and circulating in the same sense as that of the current.
Inside the solenoid the vector potential is given by equation (\ref{potcirc}) with $B=\mu_0 n I$
($n$ number of windings per unit length, $I$ current).
Outside the solenoid, the vector potential can be calculated directly from the sources \cite{rou} or  by using the {\em curl} (Stokes) theorem:
\begin{eqnarray}\label{modulo}
    A&=&\frac{\mu_0 n I a^2}{2r} \nonumber\\
       &&\\
        A_x&=&-\frac{\mu_0 n I a^2}{2r^2}y,\:\:A_y=\frac{\mu_0 n I a^2}{2r^2} x,\:\:A_z=0\nonumber
\end{eqnarray}
where $a$ the radius of the solenoid.
\subsection{Outside an ideal linear solenoid\label{outside}}
\begin{figure}[htb]
  \centering
  \includegraphics[width=4cm]{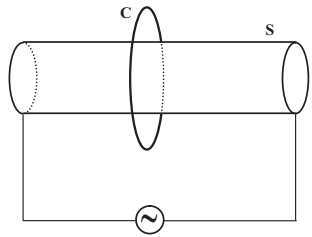}\\
  \caption{$S$ is a long solenoid in which a low frequency alternating current flows. In the metallic ring $C$, centered on the solenoid axis and lying in the median plane perpendicular to it, an alternating $emf$ is induced.}\label{ml}
\end{figure}

As recalled above, the role of vector potential in electromagnetic induction phenomena has been recently discussed in connection with the Maxwell~-~Lodge effect  \cite{rou} \cite{matteucci}  (fig. \ref{ml}).
Of course,   the flux rule predicts the correct value of the $emf$ induced  in the ring $C$. This is explicitly acknowledged by Iencinella and Matteucci \cite{matteucci}.
On the other hand, the general law (\ref{leggegen}), predicts that the $emf$ induced in the ring $C$ is given by:
\begin{equation}\label{ring}
    \mathcal E= -\oint_C \frac{\partial \vec A}{\partial t}\cdot \vec dl
\end{equation}
 Both \cite{rou} and \cite{matteucci} hold, on the basis of equation (\ref{ring}) (though not considered as a particular case of the general law (\ref{leggegen})), that the induced $emf$ in $C$ is  a local phenomenon due to the time variation of the vector potential. {\em The experiment by Blondel, that rules out any causal effect of the time variation of the magnetic flux,  proves that this interpretation is correct}. It is worth adding that Rousseaux at al. \cite{rou} treat in detail the fact that, since the solenoid is finite, the magnetic field outside is not null. Without entering in details, we only stress that the contribution of the stray magnetic field to the induced $emf$ in the ring  is opposite in sign to that calculated for an ideal solenoid \footnote{This can be seen at once by applying the flux rule: this rule also suggests that, for minimizing the effect of the stray field, the area enclosed by the ring  outside the solenoid must be as small as possible.}.
 \subsection{Inside an ideal linear solenoid}
Equations (\ref{potcirc}) and (\ref{landau1}) for the vector potential  yield different values of the electric field $-\partial \vec A/\partial t$. According to the procedure illustrated above, the ``good'' vector potential is  given by (\ref{potcirc}).
Furthermore, it is clear that (\ref{landau1}) cannot be good starting vector potentials because they yield electric fields that depend on the axis choice.
 Of course, from (\ref{potcirc}) we can get (\ref{landau1}) by using, as a gauge function, $\psi=\pm Bxy/2=\pm\mu_0nIxy/2$. The new vector potentials (\ref{landau1}) and the new scalar potential $-\partial \psi/\partial t$ yield the correct fields; however, the presence of a scalar potential different from a constant (zero) is physically meaningless since it is not justified by the sources (the current in the solenoid). The new pair of potentials constitutes only a mathematical and superfluous device for calculating the  fields: it can be dropped without diminishing the predictive power of the theory.
 \section{What's the meaning of ``physical meaning''?\label{what}}
 Looking at the history of vector potential, one should try to understand if and  how the concept of ``physical meaning'' has changed in time or in passing from a physicist to another. However, this is not the scope of the present paper. Instead, we shall try to find out a definition of ``physical meaning'' that could be applied in all situations.
  Generally speaking, it seems that, explicitly or implicity,  a ``physical meaning'' is attributed to a theoretical term if this term describes a physical quantity that can be measured.
 However, this apparently sound definition fails to be an acceptable one.
 \par
It is well known that Maxwell obtained the coefficient of rigidity of the Aether starting from the measured amount of energy coming from the Sun and falling on the Earth's surface \cite{maxet}.
  We can proceed in a similar way with the vector potential.
  For a plane wave,  the Poynting vector, averaged over a period, is given by:
      \begin{equation}\label{poinmedio}
    <S>=     \frac{\varepsilon_0 c}{2}\omega^2 A^2_0
   \end{equation}
    where   $A_0$ is
    the amplitude of the  vector potential and $\omega$ its angular frequency.
    Therefore, a measurement of the intensity and frequency of the wave yields $A_0$.
  However, as the  measurement by Maxwell of the coefficient of rigidity of the Aether does not assure a physical meaning to the Aether, so the  measurement of the amplitude of the vector potential does not assure a physical meaning to the vector potential. After all, the vector potential might be, as the Aether, a theoretical term  that can be dismissed.
  These considerations lead us to the crucial point: the physical meaning of a theoretical term relies, primarily, on theoretical grounds. We suggest that a theoretical term has a physical meaning if:
  \begin{enumerate}
       \item  [(C1)]    its elimination reduces the predictions~-~experimentally testable~-~of a theory \footnote{This criterion is a condensed version of one by Hertz's: ``I have further endeavoured in the exposition to limit as far as possible the number of those conceptions which are arbitrarily introduced by us, and only to admit such elements as cannot be removed or altered without at the same time altering possible experimental results'' \cite[page 28]{hertzprin}. Clearly, this criterion has been of basic relevance  for Hertz's rejection of the vector potential.};
           \par
           or, in a weaker sense, if
           \par
       \item [(C2)] its elimination reduces the descriptive proficiency of a theory.
                  \end{enumerate}
Electromagnetic potentials satisfy both criteria.
     As for (C1),
      the potentials allow a local and  Lorentz covariant description of  electromagnetic induction phenomena, impossible in terms of the fields \cite{epl};
      as for (C2), the potentials lead in a transparent and `spontaneous' way to a space~-~time formulation of electromagnetism.
      \par
      Criteria (C1) and (C2) can, of course, be  applied also to  the  potentials derived from the pair uniquely determined by the procedure outlined in section \ref{final}.
       These  pairs of potentials can be dropped without diminishing the predictive power of the theory; it is debatable if their dropping diminishes the descriptive proficiency of the theory (gauge invariance) \cite{jack} \cite{rou2} \cite{roubro}.
       \par
   The above criteria can be fruitfully used in dealing with a great variety of issues. Of course, their application, being theory~-~dependent, yields results that depends on time. In Maxwell's times, for instance, the Aether concept had, according to our definition, a ``physical meaning''; today, since our experimentally corroborated theories do not use the concept of Aether, this concept cannot have any ``physical meaning''.
    As significant, contemporary, examples, we shall consider only two cases: the concept of space~-~time and the wave function. The formulation of electromagnetism in terms of Minkowski space~-~time does not enlarge the predictive power of the theory (no more experimentally testable predictions): so, while space~-~time does not fulfill criterion (C1), it does criterion (C2). Of course, in general relativity, space~-~time fulfills also criterion (C1): this remind us that the application of the above criteria must take into account the entire theoretical framework.
    As for the wave function, it is  said that, though it is not a measurable quantity, it is physically meaningful because $\psi\psi^* dV$ yields the probability of finding the particle in the volume element $dV$. Also in this case, the physical meaning is attributed only in correspondence of a possible measurement. Instead, and more reasonably, the physical meaning of the wave function is simply due, according to criterion (C1), to the fact that its elimination  reduces to zero the predictive capacity of the theory.
    \par
      Up to now, we have escaped the problems connected with {\em reality}. This issue cannot be investigated here, for obvious reasons of complexity (of the issue) and brevity (of this paper). The author of the present paper has dealt with some of these arguments elsewhere \cite{ncreal}. We can only stress that the ``physical meaning'' of a theoretical term should be independent, at least at a first level, from ontological issues. The soundness of this statement can be illustrated by the following example.
    While we can setup an assertion about the {\em plausible} existence of the electron \footnote{``In the World there {\em is} a {\em quid} which has properties
that correspond to the properties attributed by our theory to the
`electron' and  this {\em quid} behaves in accordance with the
laws of our theory and with properties that are {\em described} by
the measured values of the physical quantities  that our theory
attributes to the `electron' ''. We can  convene that the statement
`the electron exists in the World' is  a {\em
shorthand} of the previous one.}, a similar assertion can hardly  be held (without  severely restricting specifications) for visible electromagnetic waves. As a matter of fact, the description of light in terms of electromagnetic waves yields predictions in conformity with experiment only when the number of photons is high enough. From this point of view, particularly illuminating are the interference experiments with low light intensities, or, better, with single photons \cite{single}.
    With reference to this last case, we see that the wave description yields the observed fringes intensity when the number of used photons is high enough. In these experiments, light {\em is} there and manifests itself as  spots on the camera's $CCD $; on the other end, we can hardly say that an electromagnetic wave {\em is} there. However, when a certain number of spots have been detected by the camera, we can describe the observed integrated intensity in terms of electromagnetic waves. In these circumstances, ontological assertions about electromagnetic waves are, at least, questionable. Nevertheless, we are unwilling to deny a ``physical meaning''  to electromagnetic waves in this case.
     Therefore, it seems that the ontological level may add further (plausible) significance to the ``physical meaning'' of a theoretical term, without, however, canceling those conveyed, more safely, by  criteria (C1) and (C2) above.
                        \section{Conclusions}
 A forgotten experiment by Blondel \cite{blondel} proves, as held on the basis of theoretical arguments in \cite{epl}, that the time variation of the magnetic flux is not the cause of the induced $emf$; therefore, as implied by the  general law of electromagnetic induction (\ref{leggegen}), the physical agent is the vector potential through the term $-\partial\vec A/\partial t$ (when the induced circuit is at rest). \par
 The ``good electromagnetic potentials'' are given by the Lorenz condition and by the retarded formulas, i.e. they are {\em uniquely  determined} by the sources. Any other couple of potentials obtained by an appropriate gauge function are only mathematical expressions which allow the calculation of the fields but are not physically related to the sources.
  \par
 The physical meaning of a theoretical term relies, primarily, on theoretical grounds: a theoretical term has physical meaning if it cannot be withdrawn without reducing the predictive power of a theory or, in a weaker sense, if it cannot be withdrawn without reducing the descriptive proficiency of a theory.
\vskip4mm\par\noindent
\small
{\bf Acknowledgements.}
I am deeply indebted to Paolantonio Marazzini for having shown me Blondel's paper.  Friendly thanks to Luigi Cattaneo and  Paolantonio Marazzini for illuminating discussions on Blondel's  paper and to Giancarlo Campagnoli for a critical reading of the first manuscript.

\end{document}